\documentclass[sigconf]{acmart}

\usepackage{booktabs}
\usepackage[normalem]{ulem}
\usepackage{amsmath}
\usepackage{pifont}
\usepackage{xfrac}
\usepackage[ruled,vlined,linesnumbered]{algorithm2e}
\usepackage{algpseudocode}
\usepackage{url}
\usepackage{comment}
\usepackage{color}
\usepackage{fancyvrb}
\usepackage{balance}
\usepackage{listings}
\usepackage{multirow}
\usepackage{enumitem}
\usepackage{mathtools}
\usepackage{graphicx}
\usepackage{pbox}
\usepackage{makecell}
\usepackage{rotating}
\usepackage{adjustbox}
\usepackage{amsthm}
\usepackage{mathtools}
\usepackage{hyperref}
\usepackage{titlesec}
\usepackage{bm}
\usepackage{multirow}
\theoremstyle{definition}

\usepackage{subfig}

\newcommand{\tool}[0]{\mbox{\textsc{DJXPerf}}}
\newcommand{\red}[1]{\textcolor{red}{#1}}

\newcommand{\milind}[1]{\textcolor{red}{milind:#1}}

\AtBeginDocument{%
  \providecommand\BibTeX{{%
    \normalfont B\kern-0.5em{\scshape i\kern-0.25em b}\kern-0.8em\TeX}}}

\setcopyright{acmcopyright}
\copyrightyear{2018}
\acmYear{2018}
\acmDOI{10.1145/1122445.1122456}

\acmConference[Woodstock '18]{Woodstock '18: ACM Symposium on Neural
  Gaze Detection}{June 03--05, 2018}{Woodstock, NY}
\acmBooktitle{Woodstock '18: ACM Symposium on Neural Gaze Detection,
  June 03--05, 2018, Woodstock, NY}
\acmPrice{15.00}
\acmISBN{978-1-4503-XXXX-X/18/06}

\begin{document}

\title{\tool{}: Identifying Memory Inefficiencies via Object-centric Profiling for Java}

%\author{Anonymous Author(s)}
\author{Bolun Li}
\affiliation{%
  \institution{North Carolina State University}
  }
\email{bli35@ncsu.edu}

\author{Pengfei Su}
\affiliation{%
  \institution{University of California, Merced}
  }
\email{psu9@ucmerced.edu}

\author{Milind Chabbi}
\affiliation{%
  \institution{Scalable Machines Research}
  }
\email{milind@scalablemachines.org}

\author{Shuyin Jiao}
\affiliation{%
  \institution{North Carolina State University}
  }
\email{sjiao2@ncsu.edu}

\author{Xu Liu}
\affiliation{%
  \institution{North Carolina State University}
  }
\email{xliu88@ncsu.edu}

\begin{abstract}
Java is the ``go-to'' programming language choice for developing scalable enterprise cloud applications.
In such systems, even a few percent CPU time savings can offer a significant competitive advantage and cost saving.
Although performance tools abound in Java, those that focus on the data locality in the memory hierarchy are rare.
%Managed languages employ various abstractions, runtime support, just-in-time compilation, and garbage collection, which hide important execution details from the plain source code. 
%Maintaining the locality of references in a CPU's memory hierarchy is well-known and mastered to achieve high performance in natively compiled code such as C and C++.  
%Although performance tools abound in Java, there are few studies focused on memory hierarchy related  in managed languages such as Java where a few percent CPU time savings can offer a significant competitive advantage and save cost.

In this paper, we present \tool{}, a lightweight, object-centric memory profiler for Java, which associates memory-hierarchy performance metrics  (e.g., cache/TLB misses) with Java objects.
\tool{} uses statistical sampling of hardware performance monitoring counters to attribute metrics to not only source code locations but also Java objects.
\tool{} presents Java object allocation contexts combined with their usage contexts and presents them ordered by the poor locality behaviors.
\tool{}'s performance measurement, object attribution, and presentation techniques guide optimizing object allocation, layout, and access patterns. 
\tool{} incurs only  $\sim$8\% runtime overhead and  $\sim$5\% memory overhead on average, requiring no modifications to hardware, OS, Java virtual machine, or application source code, which makes it attractive to use in production.
Guided by \tool{}, we study and optimize a number of Java and Scala programs, including well-known benchmarks and real-world applications, and demonstrate significant speedups. 
\end{abstract}




\maketitle

\section{Introduction}

Java is the ``go-to'' programming language choice for developing scalable enterprise cloud applications~\cite{javagoto,twitterjava,twitterjava2,hotjava,googlejava,netflixjava, uberava}.
Performance is critical in such Java programs running on a distributed system comprised of thousands of hosts; on such systems, saving CPU time even by a few percentages offers both competitive advantages (lower latency) and cost savings.
Tools abound in Java for ``hotspot'' performance analysis that can bubble-up code contexts where the execution spends the most time and are also used by developers for tuning performance.
However, once such ``low hanging targets'' are optimized, identifying further optimization opportunities is not easy.
Java, like other managed languages, employs various abstractions, runtime support, just-in-time compilation, and garbage collection, which hide important execution details from the plain source code. 
%Maintaining the locality of references in a CPU's memory hierarchy is well-known and mastered to achieve high performance in natively compiled code such as C and C++.  

In modern computer systems, compute is ``free'' but memory accesses cost dearly.
Long-latency memory accesses are a major cause of execution stalls in modern general-purpose CPUs.
CPU's memory hierarchy (different levels of caches) offers a means to reduce average memory access latency by staging data into caches and repeatedly accessing before evicting. 
Access patterns that reuse previously fetched data are said to exhibit good data locality. 
There are traditionally two types of data locality: {\em spatial} and {\em temporal}. 
An access pattern exhibits spatial locality when it accesses a memory location and then accesses nearby locations soon afterward. 
An access pattern exhibits temporal locality when it accesses the same memory  multiple times.
Programs that do not exploit these features are said to lack spatial or temporal locality.

Maintaining the locality of references in a CPU's memory hierarchy is well-known and mastered to achieve high performance in natively compiled code such as C and C++.  
Besides the traditional locality problems, garbage collected languages such as Java expose another unique locality issue --- memory bloat~\cite{Reusable}. 
Memory bloat occurs by allocating (and initializing) many objects whose lifetimes do not overlap.
For example, allocating objects in a loop where the lifetime of the object is only the scope of the loop body.
Since the garbage collection happens sometime later in the future, the memory consumption spikes, which results in a higher memory footprint and suboptimal cache utilization.
Memory bloat can be seen as a case of lack of both spatial and temporal locality because accessing a large number of independent objects results in accessing disparate cache lines and little or no reusing of a previously accessed cache line(s). 

\begin{figure}[t]
\centering
\includegraphics[width=0.45\textwidth]{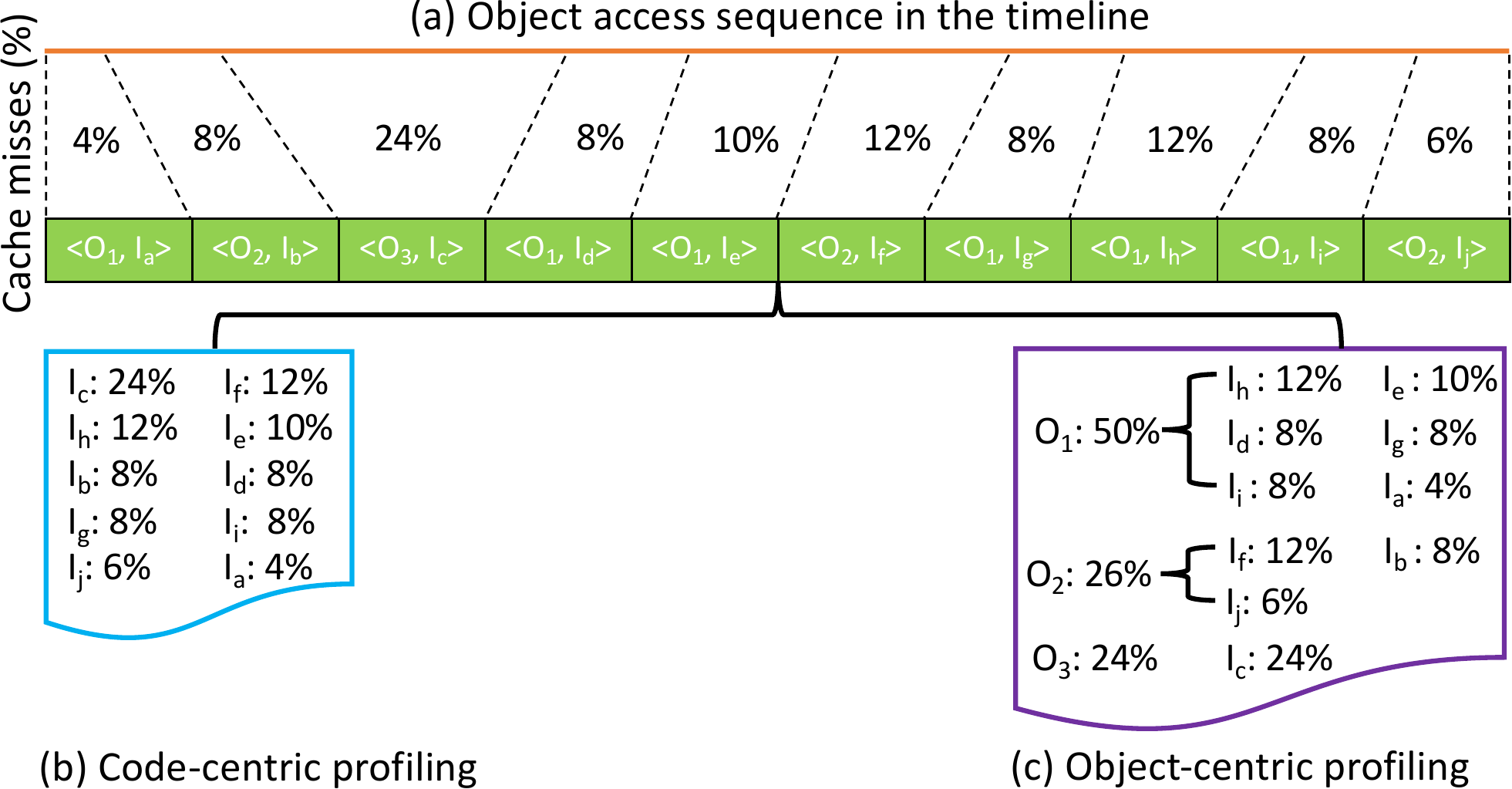}
\caption{Code-centric vs. object-centric profiling. $O$ denotes an object and $I$ denotes a memory access instruction; tuple $\langle O_m, I_n\rangle$ denotes that instruction $I_n$ accesses object $O_m$.}
\label{code-centric object-centric}
\vspace{-0.3in}
\end{figure}

Exploring data locality in native languages has been under investigation for decades. 
There exist a number of tools~\cite{ibs-cgo, ibs-pact, ibs-ispass, ibs-sc, DBLP:conf/sigmetrics/BergH05, SLO1, SLO2, reuse-chen, Liu:2015:STI:2807591.2807648} to measure the data locality using various metrics. 
Software metrics such as reuse distances~\cite{Zhong:2008, time-reuse, tree-compression, reuse-chen,macpo}, memory footprints~\cite{footprints, chunking}, and cache miss ratio curves~\cite{SLO1, CaBetacaval:2003:ECM:782814.782836}, which are derived from memory access traces, quantify the locality independent of architectures. 
In contrast, hardware metrics, such as cache/TLB misses, collected with hardware performance monitoring units (PMU) during program execution, quantify data locality on a given architecture. 
Attributing PMU metrics to source code is a straightforward way to demonstrate code regions that incur high access latencies; we call it as \emph{code-centric profiling}.

In object-oriented systems, delivering profiles centered around objects is highly desirable. 
An object may have accesses to it scattered in many places, and each code location may contribute a small fraction to the overall data-access problem.
Bridging the memory-hierarchy latency metrics with runtime data objects requires more involved measurement and attribution techniques, which is the focus of our work; we call it as \emph{object-centric profiling}. Note that \emph{object-centric profiling} not only shows the objects subject to high aggregate access latencies but also pinpoints code locations ordered by their contribution to the overall latency to the object under question.

Figure~\ref{code-centric object-centric} illustrates the difference between code-centric profiling and object-centric profiling. The code-centric profiling associates the cache miss metric with memory accesses, showing that access $I_c$ accounts for most cache misses during program execution. In contrast, the object-centric profiling aggregates the cache miss metric from different accesses that touch the same object to present a unified view. Guided by the object-centric profiling, we find that object $O_1$ accounts for most cache misses. However, its accesses are scattered across multiple instructions, and each individual access is less significant than the access to object $O_3$. %Thus, object-centric profiling reports object $O_1$ for optimization, instead of object $O_3$ that tops the code-centric profiling. 
Thus, instead of checking individual accesses, one can apply various optimization to the allocation of the $O_1$ object or its data layout. 
%The code allocates three objects, {\tt idx}, {\tt tmp}, and {\tt ph} for computation. The code-centric profiling pinpoints that lines 9 and 11 account for all L1 cache misses.
%The object-centric profiling  further associates metrics with individual objects, indicating that {\tt idx} and {\tt tmp} account for most cache misses.  

Collecting object-centric profiles for Java has unique challenges posed by managed runtimes. 
First, just-in-time compilation and interpretation used in JVM disjoin the  program source code and its execution behaviors. 
Second, automatic runtime memory management (i.e., garbage collection) further impedes understanding memory performance of Java and other languages based on JVM, such as Scala~\cite{scala-www} and Clojure~\cite{clojure-www}. 

%\milind{For people new to object-centric profiling and also for those who are fixated on code profiling, we need to show (on the first page) a picture which shows profiles ordered by allocation sites and underneath each allocation site, show the cache misses (order by high to low) at different code locations. This figure will immediately attract the readers and will help them distinguish this tool from the rest. }

Since developers in native languages have explicit knowledge and understanding of objects and their lifetimes, they pay more attention to objects and their locality; Java developers, on the other hand, lack the precise knowledge of object lifetimes and their influence on locality.
Surprisingly, Java lacks any object-centric performance attribution tool, which can pinpoint objects subject to serious latency problems. 
Tools such as Linux perf~\cite{perf} and Intel VTune~\cite{Intel:VTune} exploit hardware performance monitoring and attribute cache miss metrics to code, such as Java methods, loops, or source locations, but they do not attribute metrics to Java objects. As shown in Figure~\ref{code-centric object-centric}, these tools only attribute metrics to problematic code lines; without object-level information, they cannot tell which objects are problematic and deserve optimization.
A few tools such as~\cite{Fourtrends, XuThesis, Softwarebloat} instrument Java byte code to identify problematic objects. However, these tools suffer from high overhead and lack real execution performance metrics available from the hardware. 
Often times, the optimization lacking the quantification from the underlying hardware can yield trivial or negative speedups~\cite{ibs-ispass}.

Most of today's shared memory multiprocessor systems support non-uniform memory access (NUMA), where the loss of data locality not only pervasively exists within a single CPU processor (aka a NUMA socket), but also across CPU processors.
NUMA characteristics introduce the explicit NUMA property of the shared memory systems, where differences in local and remote memory access latencies can be up to two orders of magnitude. Shared memory applications with transparent data distributions across all nodes often incur high overheads due to excessive remote memory accesses. Minimizing the number of remote accesses is crucial for high performance, and this, in turn, usually requires a suitable, application-specific data distribution. In general, choosing an appropriate data distribution remains a challenge. We can attribute metrics to individual objects with the object-centric idea, then identifying the objects that suffer from severe remote accesses.

In this paper, we describe \tool{}, a lightweight object-centric profiler for Java programs. 
\tool{}  complements existing Java profilers by collecting memory-related performance metrics from hardware PMUs and attributing them with Java objects. \tool{}  provides unique insights into Java's memory locality issues.
 In the rest of this section, we show two motivating examples, the contribution of this paper and the paper organization.
 
\subsection{Motivating Examples}
\label{motivation}

In this section, we motivate the importance of combining object-level information and PMU metrics for locality optimization. 
Listing~\ref{motivation batik} and~\ref{motivation lusearch} show two problematic code snippets suffering from memory bloat, which are respectively from {\tt batik} and {\tt lusearch}, both from {\tt Dacapo-9.12}~\cite{dacapo}. We run them with the default large inputs using $48$ threads.

In Listing~\ref{motivation batik}, the object allocation site at line 5 creates an array of {\tt float} objects in the method {\tt makeRoom}, which is part of the  class {\tt ExtendedGeneralPath}. This allocation site is repeatedly invoked $2478$ times, resulting in memory bloat.
For optimization, one can move the array allocation outside  of the loop that encloses the method {\tt makeRoom} and replace it with a static object array, aka the singleton pattern. This optimization addresses the memory bloat and yields a $(1.15\pm0.03)\times$ speedup.

In Listing~\ref{motivation lusearch}, the memory bloat occurs at line 3, which repeatedly allocates the object {\tt collector} $15179$ times. This object is passed as an input parameter to the method {\tt search} and used in many places in that method. 
One can also apply the singleton pattern by hoisting the allocation site outside of the loop enclosing the method, and declaring it as a static object. While this optimization addresses the memory bloat, it does not bring any noticeable speedup.
%\milind{I think, the concern from previous reviews was that the code-centric profiler would not bubble up the source lines in Listing2 (because they will not encounter a lot of cache misses) and hence what is the motivation for object centric over code centric? If we can show that one of the code-centric profiler incorrectly bubbles up the code in Listing 2 (and why) then it will be more motivating.}
%\milind{A possibly different way of motivating would be to use one of the examples Xu used in his PhD work which showed that an object layout is problematic and its high latency accesses are sprayed all over the code; changing the object layout results in fixing the it in all places.}
\begin{figure}
\begin{lstlisting}[firstnumber=1,language=java,caption={The code snippet from {\tt Dacapo 9.12 batik}. Optimizing memory bloat by moving the object allocation site at line 5 outside of the loop yields a nontrivial speedup ($(1.15\pm0.03)\times$) to the entire program.},label={motivation batik}]
private void makeRoom(int numValues) {
  ...
  if ( newSize > values.length) {
    ...
@$\blacktriangleright$@  float [] nvals = new float[nlen];
    System.arraycopy(values, 0, nvals, 0, numVals);
    ... }
  ... }
\end{lstlisting}
%\vspace{-0.2in}
%\captionof{lstlisting}{The code snippet from {\tt Dacapo 9.12 batik}. Optimizing memory bloat by moving the object allocation site at line 5 outside of the loop yields a nontrivial speedup ($(1.15\pm0.03)\times$) to the entire program.}
%\label{motivation batik}
\end{figure}

\begin{figure}
\begin{lstlisting}[firstnumber=1,language=java,caption={The code snippet from {\tt Dacapo 9.12 lusearch}. Optimizing memory bloat by moving the object allocation site at line 3 outside of the loop does not bring any speedup to the entire program.},label={motivation lusearch}]
public TopDocs search(Weight weight, Filter filter, final int nDocs) {
  ...
@$\blacktriangleright$@TopDocCollector collector = new TopDocCollector(nDocs);
  search(weight, filter, collector);
  ... }
\end{lstlisting}
\vspace{-0.2in}
%\captionof{lstlisting}{The code snippet from {\tt Dacapo 9.12 lusearch}. Optimizing memory bloat by moving the object allocation site at line 3 outside of the loop does not bring any speedup to the entire program.}
%\label{motivation lusearch}
\end{figure}

The study of these two example code snippets reveals that basing the optimization only on allocation frequency (or the metrics derived from the allocation frequency~\cite{Reusable})  does not necessarily yield performance benefits.
This motivates the need for the extra locality metrics associated with the object allocation site, which we call as {\em object-centric profiling}. To be concrete, \tool{}  measures L1 cache misses\footnote{we can measure myriad other events, for example, L3 cache misses, TLB misses, etc.} with PMU on individual memory access instances and aggregates the measurement of memory accesses to the object's allocation site. 
For example, \tool{} reports that accessing the {\tt nvals} array object shown in Listing~\ref{motivation batik} accounts for 21\% of total cache misses, while accessing the {\tt collector} objects in Listing~\ref{motivation lusearch} accounts for less than 1\% of total cache misses only, which explains the different speedups obtained from the locality optimization.
%To be more specific, these cache misses could occur anywhere in the entire program, but 
\textbf{The strength of \tool{} is its ability to aggregate myriad accesses to the same object, scattered all over the program, back to the same object.}
%all such accessed associates those cache misses with the object allocation site.
We emphasize that object-centric analysis does not do away with code-centric aspect; underneath each object allocation site $\mathcal{C}$,  \tool{} provides the ability to disaggregate the code contexts contributing towards $\mathcal{C}$'s overall locality loss.
Thus, object-centric analysis with  the locality metrics associated with the object allocation sites is desired to determine whether locality optimization can yield significant speedups. 

\subsection{Paper Contributions}

In this paper, we propose \tool{}, an object-centric profiler that guides data locality optimization in Java programs. \tool{} makes the following contributions.
\begin{itemize}[leftmargin=*]
  \item \tool{} develops a novel object-centric profiling technique. It provides rich information to guide locality optimization in Java programs, which yields nontrivial speedups.
  \item \tool{} combines hardware performance monitoring units with minimal Java byte code instrumentation, which typically incurs 8\% runtime and 5\% memory overhead.
  \item \tool{} applies to unmodified Java (and languages based on JVM, e.g., Scala) applications, the off-the-shelf Java virtual machine and Linux operating system, running on commodity CPU processors, which can be directly deployed in the production environment.
  \sloppy
  \item \tool{} provides intuitive optimization guidance for developers. We evaluate \tool{} with popular Java benchmarks (Dacapo~\cite{dacapo}, NPB~\cite{npb}, Grande~\cite{grande}, SPECjvm2008~\cite{specjvm2008}, and the most recent Renaissance~\cite{Prokopec:2019:RBS:3314221.3314637}) and more than 20 real-world applications. Guided by \tool{}, we are able to obtain significant speedups by improving data locality in various Java programs.
\end{itemize}

\subsection{Paper Organization}

The paper is organized as follows. Section~\ref{relate} describes the related work and distinguishes \tool{}. Section~\ref{background} introduces some background knowledge. Section~\ref{methodology} depicts \tool{}'s methodology. Section~\ref{implementation} describes the implementation details of \tool{}. Section~\ref{evaluation} evaluates \tool{}'s accuracy and overhead. Section~\ref{case} shows some case studies. Finally, Section~\ref{conclusion} presents some conclusions.

\section{Related Work}
\label{relate}

There are numerous Java performance tools assisting developers in understanding their program behaviors, such as
profiling for execution hotspots~\cite{perf, Levon:OProfile, jprofiler-WWW, yourkit-WWW, visualvm-WWW,async-profiler-WWW, instruction-tracing} in CPU cycles or heap usage, and pinpointing redundant computation~\cite{toddler,ldoctor,Dhok:2016:DTG:2950290.2950360,DellaToffola:2015:PPY:2814270.2814290}.
These tools target orthogonal problems to \tool{}, which particularly focuses on data locality. 
Furthermore, there are many tools~\cite{ibs-cgo,ibs-ppopp,ibs-sc,Liu:2015:STI:2807591.2807648,memprof,memphis,DBLP:conf/sc/BuckH04,Roy:2016:SIF:2907294.2907308,Roy:2018:LDC:3179541.3168819} pinpointing poor locality issues in native code via OS timers or PMU-based sampling techniques, or code instrumentation. Unlike \tool{}, these tools do not work for Java applications.
In this section, we only review Java profiling techniques that are related to data locality and PMUs.

\subsection{Data Locality Analysis in Java}
Most existing Java profilers focus on memory bloat, which is one of the locality issues (aka high memory footprint) in Java. 
Mitchell \textsl{et al.}~\cite{Causes} design a mechanism to track data structures that suffer from excessive amounts of memory. Their follow-up work~\cite{Sense,Fourtrends} summarizes memory usage to uncover the costs of design decisions, which provides more intuitive guidance for code improvement. 

Xu \textsl{et al.}~\cite{Flow} develop copy profiling that detects data copies and suggests removal of allocation and propagation of useless objects. 
Their follow-up work~\cite{Containers} presents a technique that combines static and dynamic analyses to identify underutilized and overpopulated containers. They also develop a dynamic technique~\cite{Reusable} to highlight data structures that can be reused to avoid frequent object allocations. 

Nguyen and Xu~\cite{Cachetor} develop Cachetor, a value profiler for Java.
Cachetor identifies operations that keep generating identical data values and suggests memoizing the invariant values for the future usage. 
Yan \textsl{et al.}~\cite{uncovering} track object propagation by monitoring object allocation, copy, and reference operations; by constructing a propagation graph, one can identify never-used or rarely-used object allocations. 
Dufour \textsl{et al.}~\cite{Scalable, Blended} analyze the use and shape of temporary data structures based on a blended escape analysis to find excessive memory usage.
JOLT~\cite{LJOLT} uses dynamic analysis to identify object churn and performs function inlining. 

There are few studies in measuring traditional data locality in Java programs. Gu \textsl{et al.} develop ViRDA~\cite{specjvm2008-locality}, which is perhaps the most related to \tool{}. They collect memory access trace and compute reuse distance to quantify the temporal and spatial data locality in Java programs.

While these existing efforts can effectively identify some locality issues in Java, they mostly suffer from two limitations. First, they employ fine-grained byte code instrumentation, which incurs high overhead. For example, the work~\cite{Flow,Cachetor} can incur 30-200$\times$ runtime overhead. Second, they do not collect performance data from the real execution provided by the hardware; instead, they employ cache simulators. 
Without the information of the underlying hardware, optimization efforts may be misguided as shown in Section~\ref{motivation}. 

\tool{} addresses these limitations by introducing object-centric profiling technique, which is based on lightweight data collection from hardware PMUs.
\tool{} is not a replacement for existing tools; it provides complementary information to save non-fruitful optimization efforts. 

\subsection{Java Profilers Based on PMUs}
Sweeney \textsl{et al.}~\cite{pmu-java-behavior} develop a system to help interpret results obtained from PMUs when analyzing a Java application's behavior.
Cuthbertson \textsl{et al.}~\cite{pmu-runtime-components} map the instruction IP address based hardware event information to the JIT server components.
Goldshtein~\cite{profiling-JVM-production} monitors CPU bottlenecks, system I/O load and GC with perf in production JVM environments.
Hauswirth \textsl{et al.}~\cite{Vertical} introduce vertical profiling, adding software performance monitors (SPMs) to observe the behavior in the layers (VM, OS, and libraries) above the hardware. Georges \textsl{et al.}~\cite{Phase} measure the execution time for each method invocation with PMUs and study method-level phase behaviors in Java applications. Lau \textsl{et al.}~\cite{auditing} guide inline decisions in a dynamic compiler with the direct measure of CPU cycles. Eizenberg \textsl{et al.}~\cite{remix} utilize PMUs to identify false sharing in Java programs.

Unlike these existing approaches, \tool{} leverages PMUs to identify data locality in Java programs. Its usage of lightweight PMU measurement for object-centric analysis is unique among all existing Java profilers.

\subsection{NUMA Analysis in Java}
Gidra \textsl{et al.}~\cite{hal-00868012} study the scalability of throughput-oriented GCs and propose to map pages and balance GC threads across NUMA nodes. Gidra \textsl{et al.}~\cite{NumaGiC} propose local mode for NUMA machines to forbid GC threads to steal references from remote NUMA nodes so as to avoid costly cross-node memory accesses. Maria \textsl{et al.}~\cite{GCNUMA} show how to optimize three main GCs in OpenJDK, i.e., ParallelOld, ConcurrentMarkSweep, and G1 in multicore NUMA machines. Tikir \textsl{et al.}~\cite{JavaHeaps} propose NUMA-aware heap configurations for Java server applications to improve the memory performance during the GC phase. Raghavendra \textsl{et al.}~\cite{CCNUMA} propose a dynamic compiler scheme for splitting the Java code buffer on a CC-NUMA machine.

While these works can identify some memory access latency issues, they cannot pinpoint the object-level remote memory access issue. \tool{} is able to identify NUMA locality and does not depend on the GC.

\section{Background}
\label{background}
\tool{} leverages facilities available in commodity Java virtual machines (JVM) and CPU processors, which we introduce in this section.

\paragraph{\textbf{\textit{ASM Framework}}}
ASM~\cite{asm} is a Java byte code manipulation and analysis framework. ASM can modify existing classes or dynamically generate classes, supporting custom complex transformations and code analysis tools. ASM focuses on performance, with an emphasis on the low overhead, which makes it suitable for dynamic analysis. ASM can instrument object allocation (e.g., {\tt new}) and capture the object information, such as allocation size and context.

\paragraph{\textbf{\textit{Java Virtual Machine Tool Interface (JVMTI)}}}
\sloppy
JVMTI~\cite{JVMTI} is a native programming interface of the JVM, which supports developing debuggers/profilers (aka JVMTI agents) in C/C++ based native languages to inspect JVM internals. JVMTI provides a number of event callbacks to capture JVM start and end, thread creation and destruction, method loading and unloading, garbage collection epochs, to name a few. User-defined functions are subscribed to these callbacks and invoked when the associated events happen. Also, JVMTI maintains a variety of JVM internal states, such as the map from the machine code of each JITted method to byte code and source code,  and the call path for any given point during the execution. Tools based on JVMTI can query these states at any time. JVMTI is available in off-the-shelf Oracle HotSpot JVM~\cite{hotspot-jvm}.

\paragraph{\textbf{\textit{Hardware Performance Monitoring Unit (PMU)}}} 
Modern CPUs expose programmable registers (aka PMU) that count various hardware events such as memory loads, stores, and CPU cycles. 
These registers can be configured in sampling mode: when a threshold number of hardware events elapse, PMUs trigger an overflow interrupt. 
A profiler can capture the interrupt as taking a sample and attribute the metrics collected along with the sample to the execution context. PMUs are per CPU core and virtualized by the operating system (OS) for each thread.

Intel offers Precise Event-Based Sampling (PEBS)~\cite{IntelArch:PEBS:Sept09} in SandyBridge and following generations. 
PEBS provides the effective address (EA) at the time of the sample when the sample is for a memory load or store instruction. 
PEBS also reports memory-related metrics of the sampled loads/stores such as cache misses, TLB misses, memory access latency.
This facility is often referred to as address sampling --- a building block of \tool{}. 
AMD Instruction-Based Sampling~\cite{AMDIBS:07} and PowerPC Marked Events~\cite{Srinivas:2011:IBMJ-Power7} offer similar capabilities.

\paragraph{\textbf{\textit{Linux perf\_event.}}} 
Linux offers a standard interface to program and sample PMUs via the \texttt{perf\_event\_open} system call~\cite{perfevents} as well as the associated \texttt{ioctl} calls. 
The Linux kernel can deliver a signal to the specific thread whose PMU event overflows. 
The user code can extract PMU performance data and execution contexts at the signal handler.

\section{Methodology}
\label{methodology}
\begin{figure}
\centering
\includegraphics[width=0.42\textwidth]{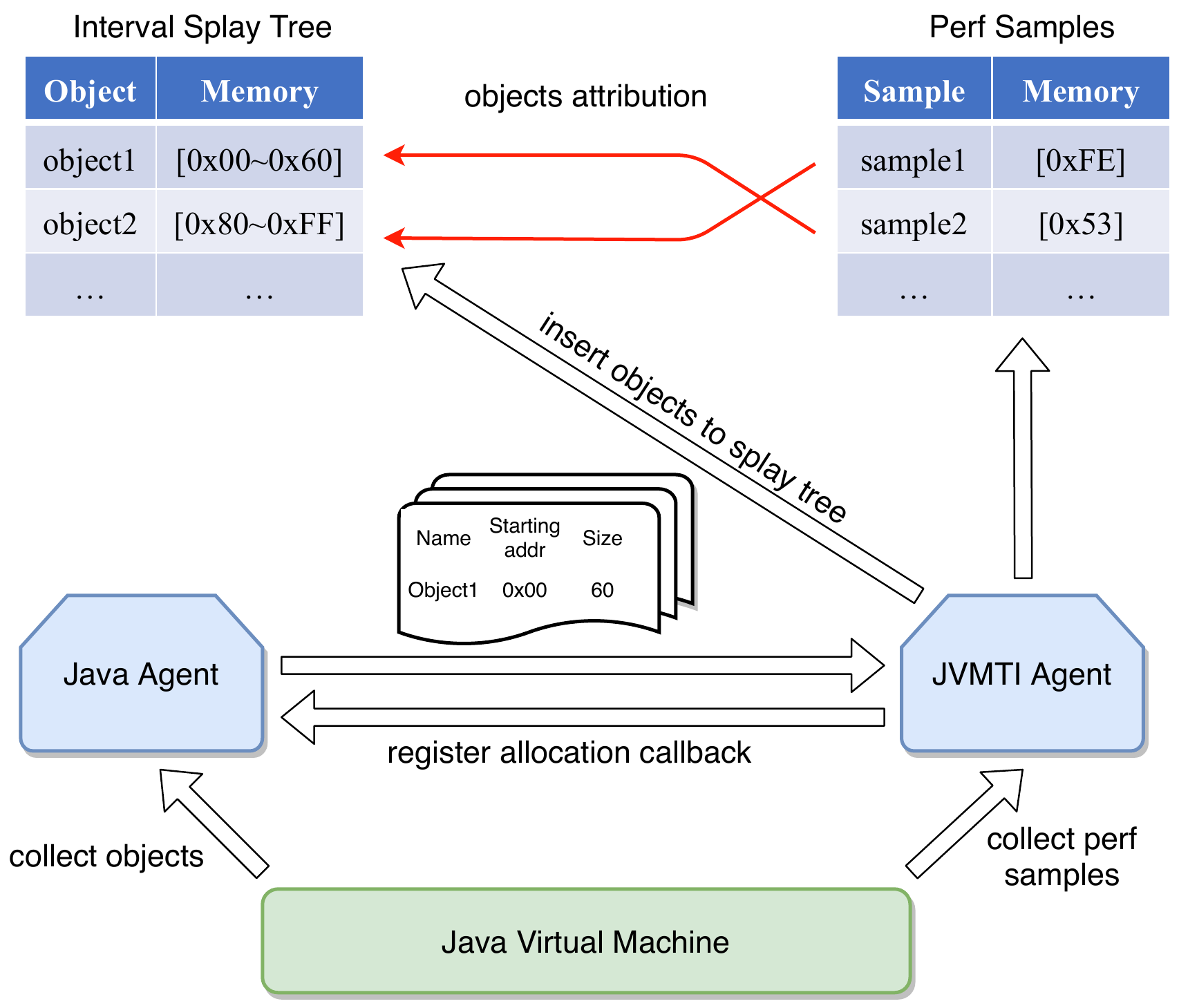}
\caption{Overview of \tool{}'s object-centric analysis.}
\vspace{-0.2in}
\label{Communication}
\end{figure}

Figure~\ref{Communication} overviews \tool{}'s object-centric profiling. \tool{} includes two agents: a Java agent and a JVMTI agent. The Java agent adds lightweight byte code instrumentation to capture object allocation information during execution, such as the allocation context and address range of objects. The JVMTI agent subscribes to Java thread creation callbacks to enable PMU to sample memory accesses. When PMU interrupts a thread with a sampled address, \tool{}  associates the address seen in the sample with the Java object enclosing that address, as shown in Figure~\ref{Communication}. In the rest of this section, we elaborate on each agent and discuss their interactions for the object-centric analysis.

\subsection{Java and JVMTI Agents}

\paragraph{\textbf{Capturing Object Addresses via A Java Agent}}
\tool{} leverages a Java agent to capture object allocation.
The Java agent is based on the ASM framework~\cite{asm}. The Java agent scans Java byte code and instruments four object allocation routines --- {\tt new}, {\tt newarray}, {\tt anewarray}, and {\tt multianewarray}. The Java agent inserts pre- and post-allocation hooks to intercept each object allocation and returns the object information (e.g., object pointer, type, and size) via user-defined callbacks. Upon each allocation callback, we follow an existing technique~\cite{object-addr} to obtain the memory range allocated for each Java object.

\paragraph{\textbf{Generating Memory Access Samples via JVMTI Agent}}
\tool{} leverages a JVMTI agent to sample and collect memory accesses. 
With the help of JVMTI, \tool{} can intercept Java thread start, where \tool{} configures PMUs to sample precise events for cache misses (e.g., {\tt MEM\_LOAD\_UOPS\_RETIRED:L1\_MISS}), TLB misses (e.g., {\tt DTLB\_LOAD\_MISSES}), or memory access latency (e.g., {\tt MEM\_TRANS\_RETIRED:LOAD\_LATENCY}). \tool{} also installs a signal handler to process PMU samples. On  thread termination, \tool{} stops PMUs and produces a profile for each thread. Besides controlling PMUs, \tool{} also utilizes the JVMTI agent to capture the calling contexts for both PMU samples and object allocations, which is described in Section~\ref{contexts}.

\subsection{Object-centric Attribution}
\label{attribution}
\paragraph{\textbf{Identifying Objects}}
Java objects are allocated on the heap. 
How to represent an object to a developer is a challenging question.
We adopt a simple and perhaps most intuitive approach that developers can identify with the allocation call path leading to the object allocation.
We represent a call path where an object $O$ was allocated with ${\mathcal P}(O)$.
An application may create multiple objects via a single allocation site, for example,  in a loop. 
In our approach, all such objects will be represented by a single call path and become indistinguishable from one another; 
we accept this trade-off since objects allocated at the same call path are likely to exhibit similar behavior.
\tool{}'s Java agent captures each allocation instance and invokes the JVMTI agent to obtain the allocation call path. As these allocation instances share the same call path, \tool{} associates the PMU metrics for any of those objects with the same call path.

%Assuming that we have a Java application which creates 10 string objects. Then, Java agent will return 10 allocation callbacks to JVMTI agent. When JVMTI agent processes these 10 string objects, it will not take these 10 string objects as 10 completely separate string objects. Instead, JVMTI agent will obtain the calling context for these 10 string objects. After that, JVMTI agent will aggregate these 10 string objects and show the allocation times for them, because the calling contexts of all of these 10 string objects are the same. 

\paragraph{\textbf{Attributing PMU Samples to Objects}}
%\milind{I am not sure this para is needed. The idea is simple, store address-range of  objects in an interval tree using Java agent, which intercepts allocation. On PMU sample, look up the sampled address in this interval tree to obtain the object.} \milind{Everything else is code details that are not necessary for understanding the idea.}
\tool{} utilizes an efficient interval splay tree~\cite{splay-tree} to maintain the memory ranges allocated for all the monitored objects. 
%A splay tree is a self-adjusting binary search tree with the additional property that frequently accessed objects have a fast lookup time. 
%It performs basic operations such as insertion, look-up and removal in O(log n) amortized time.
%Upon each object allocation, \tool{}'s Java agent invokes its JVMTI agent to query for the call path. At meanwhile, the Java agent passes the allocated memory offset and size to the JVMTI agent. The JVMTI agent maintains the $\left \langle key, value\right \rangle$ pair for each object in the splay tree, where the $key$ is the memory range $\left[start, end\right)$ allocated for the object, while the $value$ is the allocation call path for this object. The JVMTI agent inserts the $\left \langle key, value\right \rangle$ pair to the splay tree as a node.
On each PMU sample, \tool{} uses the effective address $M$ presented by the PMU to look up into the splay tree. 
The lookup for $M$ returns the object $O$ whose memory range encloses the sampled address. 
\tool{} then attributes any associated PMU metric related with the sample to ${\mathcal P}(O)$ --- the object's allocation call path.

\subsection{Object NUMA locality detection}
On each PMU sample, \tool{} uses the effective address $M$ to identify a memory page in libnuma library function {\tt numa\_move\_pages}, which simply uses the {\tt move\_pages} system call. Not only can {\tt move\_pages} move a specified page to a specified NUMA node, but also return the NUMA node where the page is currently residing. Then \tool{} is able to identify which NUMA node ($Node_1$) the current object is allocated. Still on each PMU sample, \tool{} uses {\tt PERF\_SAMPLE\_CPU} identifier, which contains the CPU number, to show which NUMA node ($Node_2$) the current object is accessing. If $Node_1$ and $Node_2$ are distinct nodes, then \tool{} reports a remote memory access for this object.

\subsection{Calling Context Determination}
\label{contexts}
%\milind{Section 4.4 should appear before talking about Obtaining Calling Contexts.}
We associate an object allocation with the full calling context (aka call path) leading to its allocation.
The full call path helps distinguish allocations by the same routine called from different code contexts.
The alternative, a flat profile, would be unable to distinguish, for example, an allocation in a common library routine called from two different user code locations.

Oracle Hotspot JVM supports two JVMTI APIs to obtain calling contexts: \texttt{GetStackTrace} and \texttt{AsyncGetCallTrace}. \texttt{GetStackTrace} requires the program to reach a safe point to collect calling context, which produces biased results~\cite{Mytkowicz:2010:EAJ:1806596.1806618,Hofer:2014:FJP:2647508.2647509}. 
Instead, \tool{} employs \texttt{AsyncGetCallTrace} to obtain calling contexts at anytime~\cite{AsyncGetCallTrace-WWW}. \texttt{AsyncGetCallTrace} accepts \texttt{u\_context} obtained at PMU interrupts or object allocations, and returns the byte code index (BCI) and method ID for each frame in the calling context. Usually, a single Java source line may translate to several byte code instructions, and the BCI can tell which byte code instruction was executed. 
Since an individual method may be JITted multiple times, the method ID helps distinguish different JITted instances of the same method. 
With the method ID, \tool{} obtains the corresponding class name and method name by querying JVM. To obtain the line number, \tool{} maintains a ``\texttt{BCI$\rightarrow$line number}'' map for each method instance via JVMTI API \texttt{GetLineNumberTable} and queries the line number on demand. %As a result, for any given BCI, \tool{} returns its line number by looking up the mapping table. 

\subsection{Interfering with Garbage Collection}
The garbage collector (GC) complicates object-centric attribution because GC implicitly reclaims memory of unused objects and  moves objects for compact memory layouts. The trigger of the GC thread is determined by JVM, which is transparent to Java applications.  Ignoring GC, \tool{} may yield incorrect object attribution. 
We assume GC reclaims or moves an object $\mathcal{O}_1$, whose memory  can be reused by another allocation, say object $\mathcal{O}_2$. 
In this case, a tool may incorrectly attribute PMU samples that touch $\mathcal{O}_2$ to $\mathcal{O}_1$. Moreover, if $\mathcal{O}_1$ is moved to a new memory location, any subsequent PMU samples of the new address will be unavailable for us to map via the original mapping maintained in the splay tree.

Handling GC is necessary in \tool{}. Unfortunately, JVM exposes limited information about GC via JVMTI: 
JVMTI only provides hooks to register callbacks on GC start and end, with no insight about the individual object behavior (i.e., reclamation and movement). %Even worse, such GC events are not general to all GC types; only stop-the-world GC can be monitored. 
We offer a solution to handle all kinds of GC in the off-the-shelf JVM.

\paragraph{\textbf{Solution for Object Movement by GC}}
Our solution is based on an important observation from the source code of OpenJDK: {\em GC moves objects using the {\tt memmove} function.} Thus, \tool{} overloads {\tt memmove} to obtain the source and destination of every moved object and update the memory ranges associated with this object in the splay tree described in Section~\ref{attribution}. However, updating the splay tree upon each {\tt memmove} invocation is costly. Instead, \tool{} creates a relocation map for each thread to record the moved objects (e.g., source as the key, destination memory addresses, and size as the value). \tool{} updates the objects in the map in a batch at the end of each GC invocation. 

To capture every GC invocation, \tool{} utilizes JVM management interface, i.e., {\tt MXBean}.  \tool{}, in the Java agent, registers GC invocation callbacks via  {\tt GARBAGE\_COLLECTION\_NOTIFICATION} event. Upon each GC completion, an {\tt MXBean} instance (i.e., {\tt GarbageCollectorMXBean}) emits a callback; \tool{} captures this callback and updates all the newly moved objects in the relocation map to the splay tree. The relocation map is reset after the update.

It is worth noting that \tool{} may not always capture all the object allocation because its attach mode may omit  some allocations (see Section~\ref{attach}). If this is the case, \tool{} directly inserts the new memory intervals for the moved objects.

\paragraph{\textbf{Solution for Object Reclamation by GC}}
\tool{} handles object reclamation by overloading the {\tt finalize} method. GC always calls the {\tt finalize} method before reclaiming memory for any object, which cleans up resources allocated to the object. \tool{} intercepts the {\tt finalize} method, obtains the memory interval reclaimed, and removes it from the splay tree.

\section{Implementation}
\label{implementation}

\begin{figure}
\centering
\includegraphics[width=0.4\textwidth]{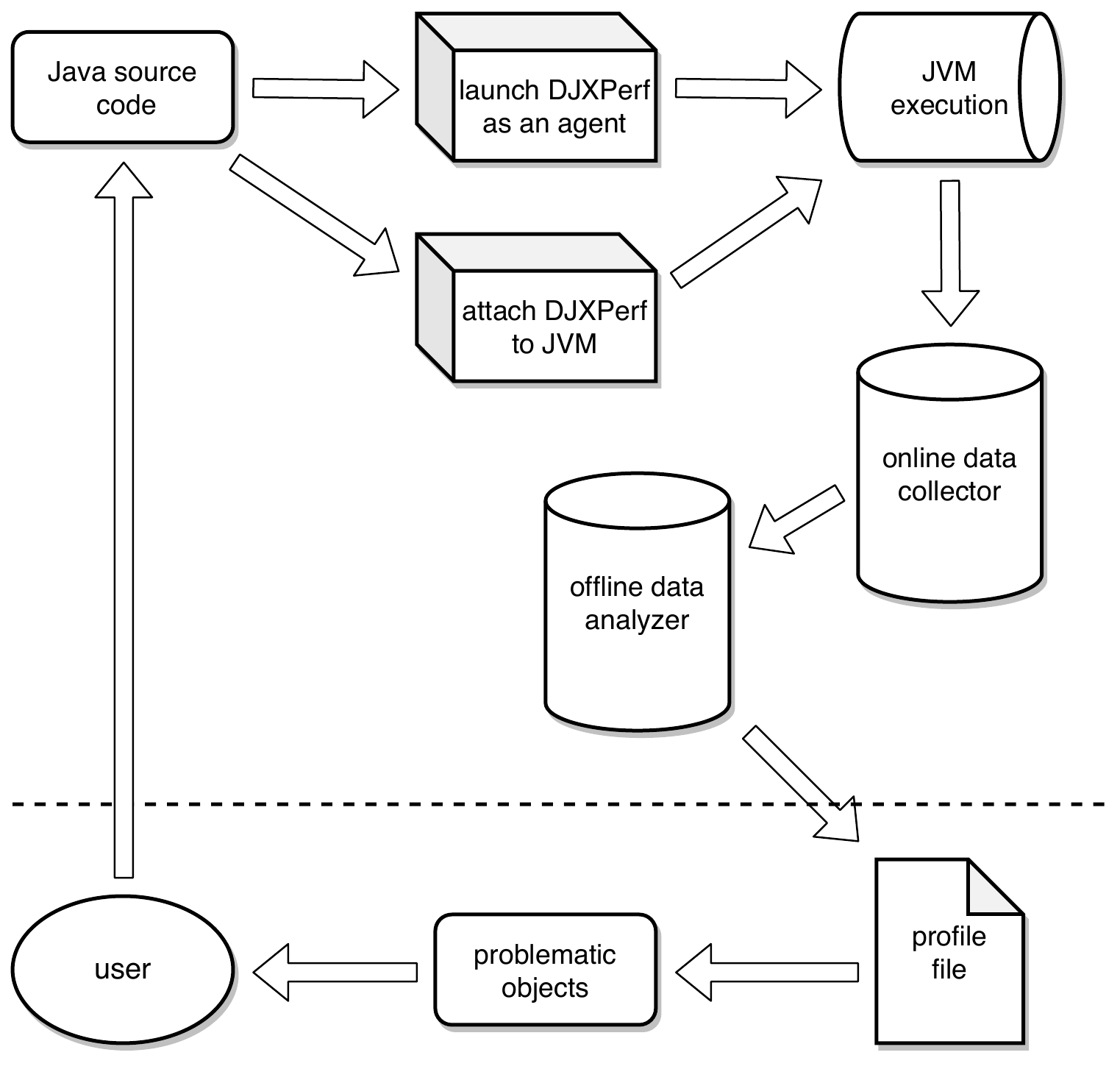}
\caption{The workflow of \tool{}.}
\vspace{-1em}
\label{workflow}
\end{figure}

\tool{} is a user-space tool with no need for any privileged system permissions.
\tool{} requires no modification to hardware, OS, JVM, or monitored applications, which makes \tool{} applicable to the production environment. 
Figure~\ref{workflow} shows the workflow of \tool{}, which consists of an online data collector and an offline data analyzer. There are two ways to enable the collector. If we need to profile Java source code as soon as the JVM starts up, we can launch \tool{} as an agent by passing JVM options. If the JVM is already started, we can attach \tool{} to this running JVM. The collector gleans the measurement via the Java and JVMTI agents and generates a profile file per thread. The analyzer then aggregates the files from different threads, sorts the metrics, and highlights the problematic objects for  investigation.

%Figure \ref{workflow} shows the workflow of \tool{}. \tool{} consists of two components: data collector and data analyzer. At the beginning of workflow, we have the source code of Java application. Then we compile it to executable binaries. After that, data collector in Java agent takes the executable binaries as input to track object information. And then, data analyzer in JVMTI agent does the object-centric analysis. Finally, the data analyzer generates a profile file. In multi-threads running environment, each thread maintains their own profile and data analyzer will aggregate them as a single profile file for viewing the problematic objects.

The implementation challenges include maintaining a low measurement overhead and scaling the analysis to many threads. In the rest of this section, we discuss how \tool{} addresses these challenges. 

\subsection{Online Collector}
\label{attach}
\tool{} supports two modes to monitor a Java program. On the one hand, \tool{} can monitor the end-to-end execution of a program by launching the tool together with the program. On the other hand, \tool{} can attach and detach to any running Java program to collect the object-centric profile for a while. This is particularly useful to monitor long-running programs such as web servers or microservices. 
 
\tool{} accepts any memory-related PMU precise events. In our implementation, \tool{} presets the event as L1 cache misses (MEM\_LOAD\_UOPS\_RETIRED:L1\_MISS). We empirically choose a sampling period to ensure \tool{} is able to collect 20-200 samples per second per thread, which has a good trade-off between runtime overhead and statistical accuracy~\cite{Tallent:2010:phd}.

To minimize the thread synchronization, \tool{} has each thread collect PMU samples independently and maintains the calling contexts of PMU samples in a compact calling context tree (CCT)~\cite{Arnold-Sweeney:1999:cct}, which merges all the common prefixes of given calling contexts. The only shared data structure between threads is the splay tree for objects because an object allocated by a thread can be accessed by other threads. \tool{} uses a spin lock to ensure the integrity of the splay tree across threads.

Another source of overhead is monitoring objects, which depends on a given Java application. \tool{} can either monitor every object allocation, or filter out objects whose sizes are smaller than a configurable value $\mathcal{S}$ from monitoring to trade off the overhead. \tool{} by default sets $\mathcal{S}$ as 1KB to pinpoint large objects that are more vulnerable to locality issues. We evaluate the impact of $\mathcal{S}$ in Section~\ref{evaluation}.

%Monitoring frequent object allocation can incur high overhead. Thus, instead of monitoring every object allocation, \tool{} can filter out objects whose sizes are smaller than a configurable value $\mathcal{S}$ from monitoring to trade off the overhead. In our experiments, \tool{} sets $\mathcal{S}$ as 1KB to 
%pinpoint large objects that are more vulnerable to suffer from locality issues. This strategy greatly reduces the overhead. For example, DaCapo chart originally has 52\% runtime slowdown and now gets reduced to 4\%.

\begin{comment}
%In this way, \tool{} has a low runtime overhead for collecting and attributing objects. We evaluate \tool{}'s overhead in Section 6.
\milind{How can we convince anybody that 1KB is the right choice? We should say that it is configurable. At a high-level, object bloat can occur with small objects also. What is the breakdown of the 52\% overhead into allocation instrumentation overhead, splay-tree lock wait, tree insertion cost, tree lookup cost? }
\milind{I have ideas here: objects allocated by one thread are likely accessed by the same thread. Hence, instead of a shared splay tree, maintain a binary tree per thread. Each thread looks up its own tree before looking up a different thread's tree (most likely it finds its addresses in its tree). I will allow concurrent readers but a  writer (allocation) blocks new readers. }
\end{comment}

\begin{comment}
\begin{figure}
\centering
\includegraphics[width=0.3\textwidth]{cct.pdf}
\caption{The calling context tree (CCT) for objects attribution.}
\label{cct}
\end{figure}
\end{comment}

\begin{figure*}[t]
\begin{center}
\subfloat[Runtime overheads.]{
\includegraphics[width=0.49\textwidth]{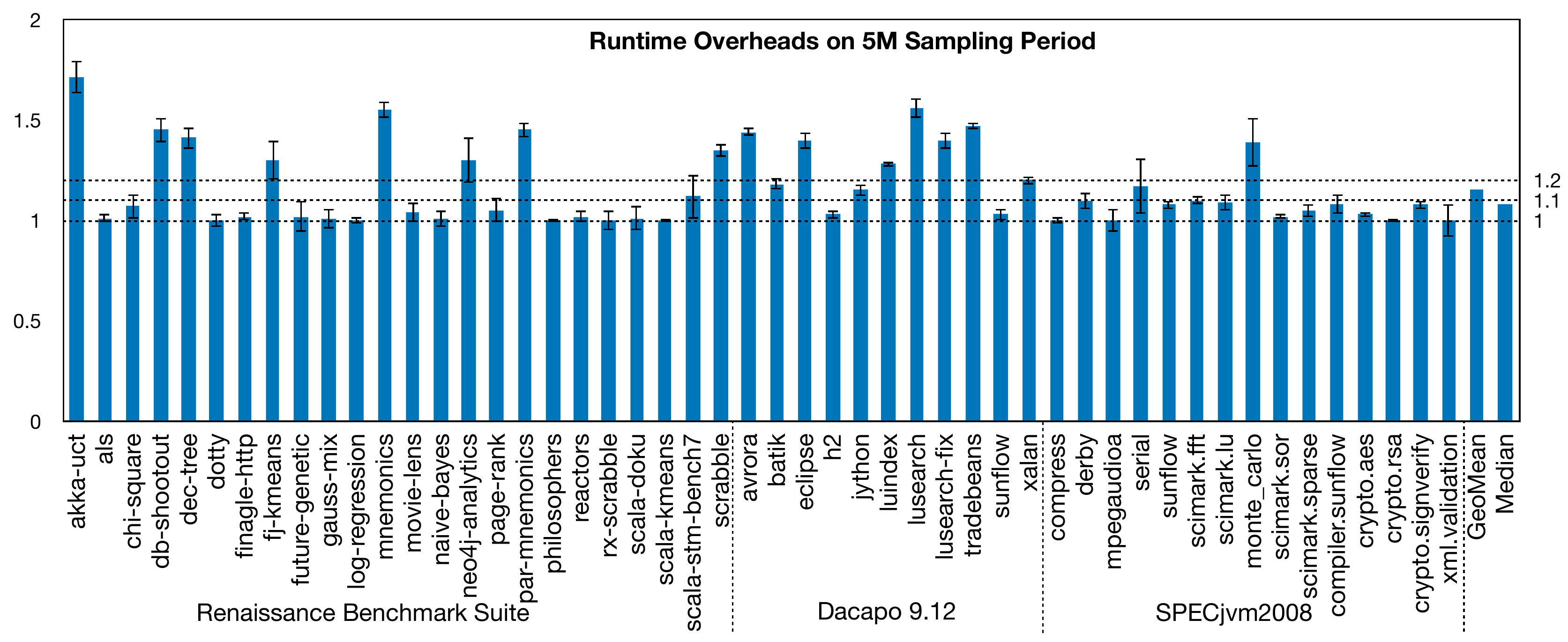}
%\vspace{-1em}
\label{fig:runtime-slowdown}
}
~
\subfloat[Memory overheads.]{
\includegraphics[width=0.49\textwidth]{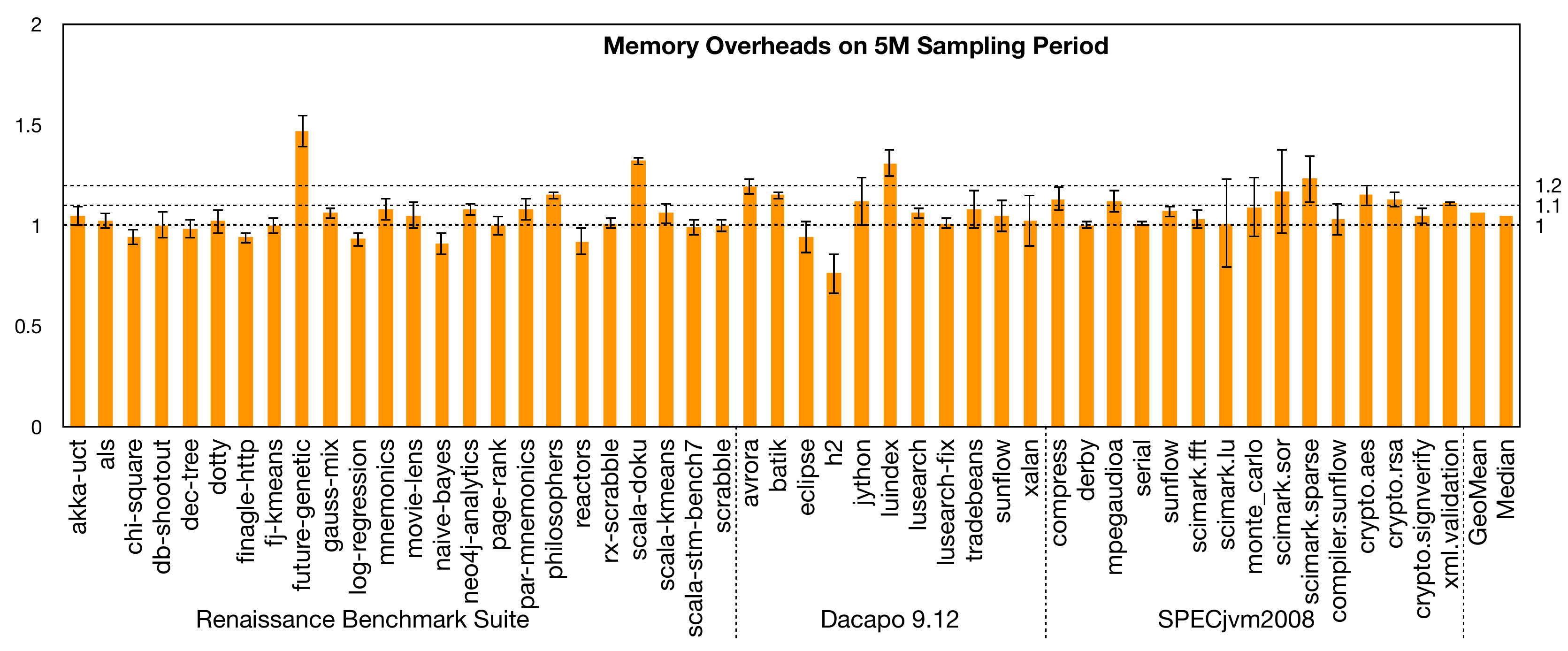}
%\vspace{-1em}
\label{fig:memory-bloat}
}
\end{center}
%\vspace{-0.1in}
\caption{\tool{}'s runtime and memory overheads in the unit of times ($\times$) on various benchmarks.}
\label{fig:overhead-5M}
\end{figure*}

\subsection{Offline Analyzer}
To generate compact profiles, which is important for scalability, \tool{}'s offline analyzer merges profiles from different threads. Object-centric profiles, organized as a CCT per thread, are amenable to coalescing. The analyzer merges CCTs in a top-down way --- from call path root to leaf. 
Call paths for individual objects and their memory accesses can be merged recursively across threads. If object allocation call paths are the same, they have coalesced even if they are from different threads. All memory accesses with their call paths to the same objects are merged as well. Metrics are also summed up when merging CCT nodes. Typically, \tool{}'s analyzer requires less than one minute to merge all profiles in our experiments. \tool{} provides a Python-based GUI to visualize the profiles for intuitive analysis.

\subsection{Discussions}
\label{threat}
\begin{comment}
\tool{} is a sampling-based tool. The sampling strategy only identifies statistically significant performance issues and ignores some insignificant ones. Also, the sampling rate should be appropriately chosen. 
A high sampling rate brings high overhead, and a low sampling rate obtains insufficient samples, which can result in over- or under-estimation. Moreover, our experiments are done on an Intel processor. 
One may not get the same speedups when running applications on a different processor from ARM, AMD, and IBM.
Finally, to solve the memory bloat issues detected by \tool{}, we avoid the repeated creations of objects by reusing the object upon request instead of creating a new one. In some cases, however, repeated object creation is used intentionally. For instance, with the security concerns, developers may prefer to create a new object instance to clear confidential information stored in the previous object instance. Thus, \tool{} produces the objects potentially for optimization, and users need to revise the code for optimization manually.
\end{comment}

\tool{} is a sampling-based dynamic profiling tool. The sampling strategy only identifies statistically significant performance issues and ignores some insignificant ones. Also, the sampling rate should be appropriately chosen. 
A high sampling rate brings high overhead, and a low sampling rate obtains insufficient samples, which can result in over- or under-estimation. 
Nevertheless, there is sufficient evidence to show that random sampling via PMUs is superior to biased sampling~\cite{Mytkowicz:2010:EAJ:1806596.1806618}.
Like other dynamic profilers \tool{}'s optimization guidance is input dependent. 
We recommend to use typical program inputs for representative profiles. Additionally, we ensure the optimization  is correct across different inputs. 
Finally, \tool{} pinpoints objects potentially for optimization, but users need to determine and apply the optimization.

%to solve the memory bloat issues detected by \tool{}, we avoid the repeated creations of objects by reusing the object upon request instead of creating a new one. In some cases, however, repeated object creation is used intentionally. For instance, with the security concerns, developers may prefer to create a new object instance to clear confidential information stored in the previous object instance. Thus, and users need to revise the code for optimization manually.

\section{Evaluation}
\label{evaluation}
We evaluate \tool{} on a 24-core Intel Xeon E5-2650 v4 (Broadwell) CPU clocked at with 2.2GHz running Linux $4.18$. 
The memory hierarchy consists of a private 32KB L1 cache, a private 256KB L2 cache, a shared 30MB L3 cache, and 256GB main memory. \tool{} works for any Oracle JDK version with the JVMTI support; in this experiment, we run all applications in Oracle Hotspot JVM with JDK 1.8.0\_161.

\paragraph{\textbf{Accuracy Analysis}}
We show that \tool{} can accurately pinpoint Java objects with the poor locality.
We evaluate \tool{}'s ability to find performance bugs on  five benchmarks that have locality issues reported by prior work~\cite{Reusable}. 
These five benchmarks are luindex, bloat, lusearch, and xalan, all from  Dacapo 2006~\cite{dacapo}, as well as SPECJbb2000~\cite{specjbb2000}. \tool{}  successfully identified all the locality issues as reported by existing tools~\cite{Reusable}. We elaborate the analysis of these benchmarks in the appendix since they have already been discussed elsewhere.

\paragraph{\textbf{Overhead Analysis}}
The runtime overhead (memory overhead) is the ratio of the runtime (peak memory usage) under monitoring with \tool{} to the runtime (peak memory usage) of the corresponding native execution. To quantify the overhead of \tool{} in both runtime and memory, we apply \tool{} to a number of well-known Java benchmark suites, such as the most recent JVM parallel benchmarks suite Renaissance~\cite{Prokopec:2019:RBS:3314221.3314637}, Dacapo 9.12~\cite{dacapo-9.12}, and SPECjvm2008~\cite{specjvm2008}. We run all benchmarks with four threads if applicable.
We run every benchmark 30 times and compute the average and error bar.
%Also, in order to show that our profiler can provide some guidances for real java development process, we have tested the profiler on many real Java applications. All the java benchmarks are run on a quad-core machine with an Intel Xeon E5-2699 2.3 GHz processor, running Linux 3.10.
%Table \ref{overhead_table} and 
Figure~\ref{fig:overhead-5M} shows the overhead when \tool{} is enabled at a sampling period of 5M for Renaissance benchmark suite~\cite{Prokopec:2019:RBS:3314221.3314637}, Dacapo 9.12~\cite{dacapo-9.12}, and SPECjvm2008~\cite{specjvm2008}. Some Renaissance and Dacapo benchmarks have higher time overhead (larger than 30\%) because they usually invoke too many allocation site callbacks (e.g., more than $400$ million times for mnemonics, par-mnemonics, scrabble, akka-uct, db-shootout, dec-tree, and neo4j-analytics). From Figure~\ref{fig:overhead-5M} we can see that \tool{} typically incurs 8\% runtime and 5\% memory overhead.
%The average time and space overhead are 13\% and 7\%, respectively.

\paragraph{\textbf{Further Discussions}}
To trade off the overhead, \tool{} allows users to set up a configurable value $\mathcal{S}$ to capture memory allocations with the size greater than $\mathcal{S}$. We also test an extreme case: setting $\mathcal{S}$ to 0, which means \tool{} captures the allocation for each object. With the evaluation on the Renaissance benchmark suite~\cite{Prokopec:2019:RBS:3314221.3314637}, \tool{} incurs a runtime overhead ranging from $1.8\times$ to  $3.6\times$ to monitor every object allocation. With further investigation, we seldom find opportunities for optimizing small-sized objects.  Thus, we believe setting $\mathcal{S}$ to 1KB is a good trade-off between the overhead and obtained insights.
One can expect \tool{} to be used in attach and detach mode on production services, where developers collect profiles from multiple instances of their services; hence, the overhead, if any, is introduced only for the short duration of measurement and the samples from multiple instances will offer a good coverage.

\begin{comment}
\begin{figure*}
\centering
\includegraphics[width=\textwidth]{overhead_plot.pdf}
\caption{\tool{}'s runtime and memory overheads in the unit of times ($\times$) on various benchmarks.}  \red{(psu: add the dividing line between GeoMean and SPECjvm2008)} \milind{Space and time overheads should be two different figures, not mixed together.}
\label{overhead_plot}
\end{figure*}
\end{comment}

\begin{table*}
\centering
\includegraphics[width=0.8\textwidth]{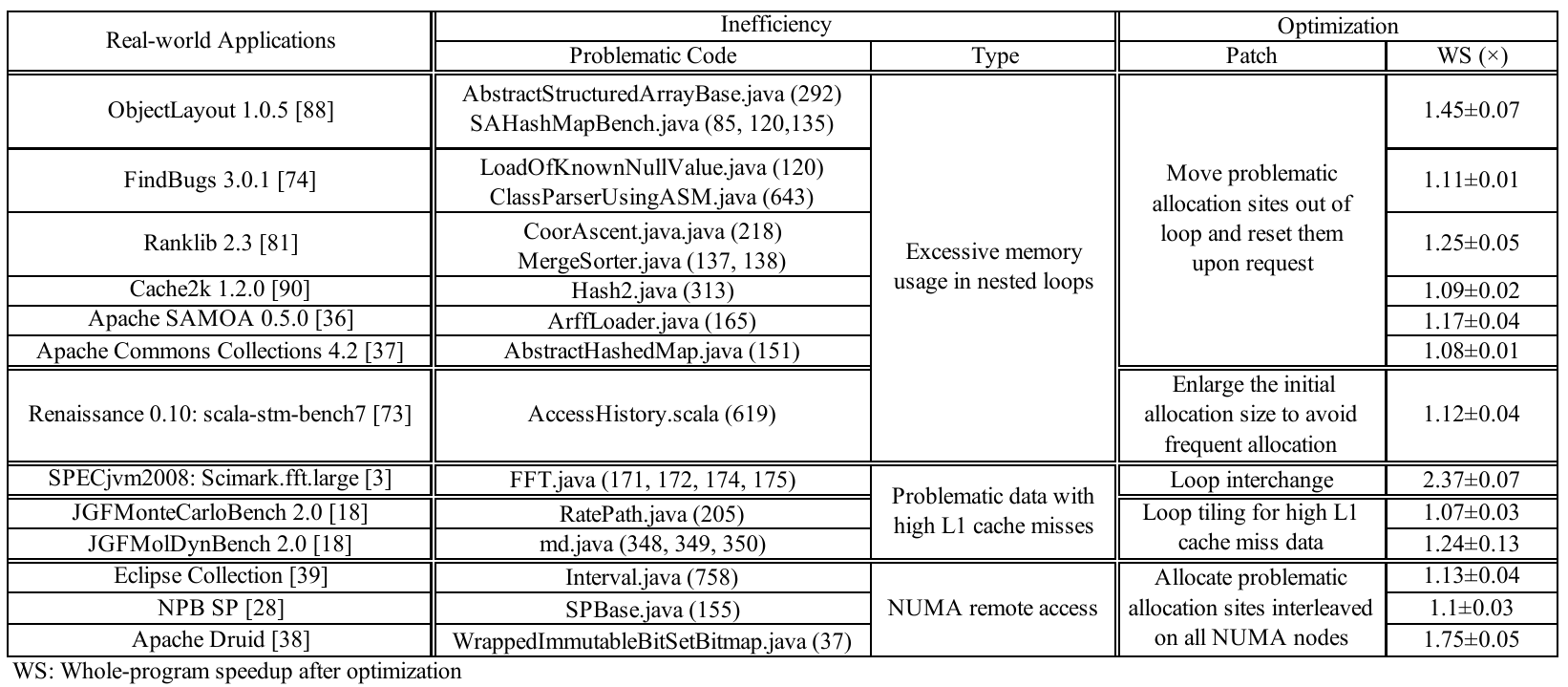}
\caption{Overview of performance optimization guided by \tool{}.}
\vspace{-1em}
\label{overview}
\end{table*}

\section{Case Studies}
\label{case}
\tool{}'s low overhead allows us to collect object-centric profiles from a variety of Java and Scala applications, such as the Renaissance benchmark suite~\cite{Prokopec:2019:RBS:3314221.3314637}, ObjectLayout~\cite{ObjectLayout}, Findbugs~\cite{FindBugs}, Ranklib~\cite{Ranklib}, cache2k~\cite{cache2k}, Apache SAMOA~\cite{SAMOA}, Apache Commons Collections~\cite{Commons}, Java Grande 2.0~\cite{grande}, to name a few. 
We run these applications with the default inputs released with these applications or the inputs that we can find to our best knowledge. To fully utilize CPU resources, we run each parallel application to saturate all CPU cores if not specified. \tool{} can pinpoint many previously not reported data locality issues and guide optimization choices. To guarantee correctness, we ensure the optimized code passes the application validation tests.
To avoid system noises, we run each application 30 times and use a 95\% confidence interval for the geometric mean speedup to report the performance improvement, according to the prior approach~\cite{inefficiencies-in-java}. Table~\ref{overview} overviews the performance improvements on several real-world Java applications guided by \tool{}. In the remaining section, we elaborate on the analysis and optimization of several applications under the guidance of \tool{} in Section~\ref{case1}-~\ref{case2}. Due to the page limit, we omit the description of other applications in Table~\ref{overview}. % in the supplementary material~\cite{supplementary}. 
Section~\ref{adversary} shows some studies on optimizing insignificant objects, which illustrate the unique usefulness of \tool{} over other tools. 

We also collected code-centric profiles using the Linux perf utility for each case study to compare with \tool{}'s profiles.
In several cases, we found it arduous to tie data accesses segregated over many code location back to the object allocation site in code-centric profiles.
\tool{} eased developer's task by showing top data objects subject to locality problems along with an ordered, hierarchical view of code locations contributing to the locality problem. 

To optimize the NUMA locality issue, we developed a Java library to access the libnuma library interfaces. As the libnuma library can be used only in native languages, we leverage Java Native Interface (JNI) to enable our Java library to access the native NUMA API. After \tool{} detected the problematic object that suffers from remote memory access, we modify the Java application allocation code for this object by calling the libnuma {\tt numa\_alloc\_interleaved} API. Then, we can allocate this problematic object interleaved on all NUMA nodes to reduce the remote accesses.

%The remaining section consists of three parts. Section~\ref{bloat} shows how \tool{} detects data locality issues due to memory bloat in Java programs.
%Section~\ref{locality} shows how \tool{} pinpoints traditional data locality issues in Java programs.
%Section~\ref{adversary} shows some studies on optimizing insignificant objects, which illustrate the unique usefulness of \tool{} over other tools. 

%\subsection{Locality Issues due to Memory Bloat}
%\label{bloat}
\subsection{ObjectLayout 1.0.5}
\label{case1}
\begin{figure}
\centering
\includegraphics[width=.49\textwidth]{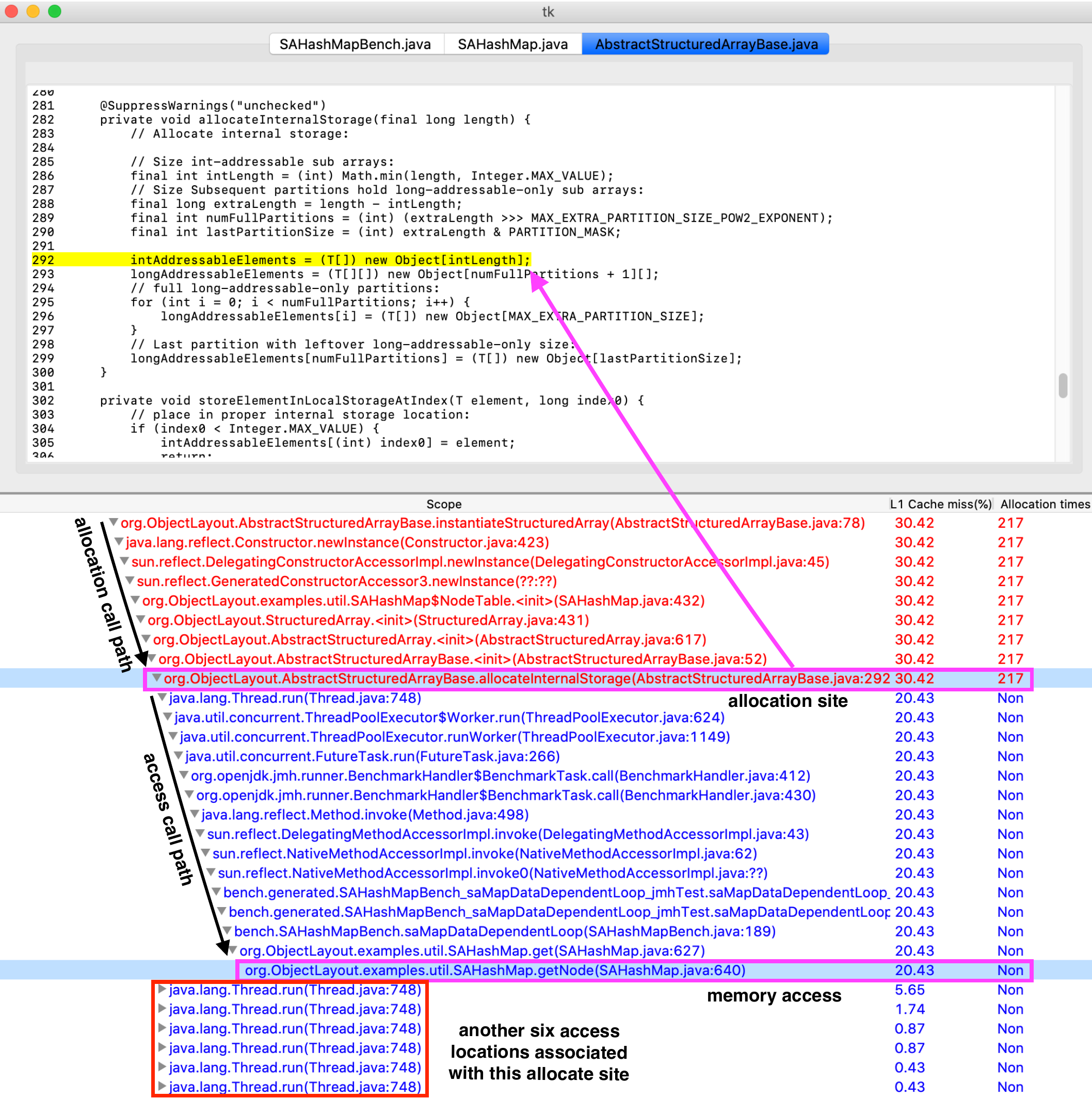}
\caption{The top-down object-centric GUI view of ObjectLayout shows a problematic object's allocation site in source code, and its allocation call path with its all access call paths.}
\vspace{-1em}
\label{gui}
\end{figure}

ObjectLayout~\cite{ObjectLayout} is a layout-optimized Java data structure package, which provides efficient data structure implementation. We run ObjectLayout with the SAHashMap input released with the package. \tool{} reports four problematic objects, which account for 84\% of cache misses in the entire program. 
Figure~\ref{gui} shows the snapshot of \tool{}'s GUI for intuitive analysis. 
The top pane of the GUI shows the Java source code; the bottom left pane shows the object allocation (in red) and accesses (in blue) in their full call paths; and the bottom right pane shows the metrics (e.g., L1 cache misses, object allocation times in this example).  

The GUI in Figure~\ref{gui} shows one problematic object, which is at line 292 of method {\tt allocateInternalStorage} in class {\tt AbstractStructuredArrayBase}. The {\tt allocateInternalStorage} method is repeatedly invoked in a loop when a new instance is created by the {\tt newInstance} method.
The bottom left pane of the GUI shows one problematic object allocation in the full call path, which is allocated 217 times in a loop. There are multiple sampled accesses associated with this allocation site and they account for 30.4\% L1 cache misses of the entire program. To save the space, we only show one access in its call path in blue (method {\tt getNode} in class {\tt SAHashMap}) that accounts for most of the cache misses. For the other accesses, we only show their call paths rooted at {\tt java.lang.Thread.run}; with no object-centric profiling, these accesses are separately presented with no aggregate view.

We further investigate the source code and find that the problematic object is array {\tt intAddressableElements}. The life cycles of different instances do not overlap, which means using singleton pattern of this object (i.e., allocating a single object instance and reusing it without creating more instances) is safe and avoids memory bloat. To apply the singleton pattern, we hoist {\tt intAddressableElements} allocation out of {\tt allocateInternalStorage} method, which is thread safe. 
We optimize  three other problematic objects with similar methods. 
Our optimization reduces total cache misses  by 76\% and yield a $(1.45\pm0.07)\times$ increase in throughput.

\begin{comment}
\begin{figure}
\begin{lstlisting}[firstnumber=279,language=java,caption={The source code related to the problematic objects' allocations in ObjectLayout.}]
static <S extends AbstractStructuredArrayBase<T>, T> S instantiateStructuredArray(...) {
  ConstructorMagic constructorMagic = getConstructorMagic();
  constructorMagic.setConstructionArgs(arrayModel);
  try {
    constructorMagic.setActive(true);
    @$\blacktriangleright$@return arrayConstructor.newInstance(args);
    //newInstance method will call to allocateInternalStorage method below
  } ...
  ...
}

private void allocateInternalStorage(final long length) {
  ...
  @$\blacktriangleright$@intAddressableElements = (T[]) new Object[intLength];
  ...
}
\end{lstlisting}
\vspace{-0.3in}
%\captionof{lstlisting}{The source code related to the problematic objects' allocations in ObjectLayout.}
\label{objectlayout1}
\end{figure}
\end{comment}

\subsection{FindBugs 3.0.1}
FindBugs is a program to find bugs in Java programs. It looks for instances of ``bug pattern'' --- code instances that are likely to be errors~\cite{FindBugs}. We run FindBugs on Java chart library 1.0.19 version as input.
\tool{} reports two objects that account for 32\% of cache misses in the entire program as shown in Listings~\ref{FindBugs1} and~\ref{FindBugs2}.
The two problematic objects {\tt buf} and {\tt IdentityHashMap} are both repeatedly allocated in loops with no overlap in lifecycles across different instances. We apply singleton pattern by hoisting the two object allocations out of the loops to avoid memory bloat. These optimizations reduce peak memory usage from 1.8GB to 0.9GB, yielding a $(1.11\pm0.01)\times$ speedup to the entire program.

\begin{figure}
\begin{lstlisting}[firstnumber=634,language=java,caption={The problematic source code highlighted by \tool{} in FindBugs.},label=FindBugs1]
public void setAppClassList(List<ClassDescriptor> appClassCollection) {
  for (ClassDescriptor appClass : allClassDescriptors) {
     ...
     @$\blacktriangleright$@XClass xclass = currentXFactory().getXClass(appClass); 
     //getXClass method will call to parse method below
  }}
public void parse(ClassInfo.Builder builder) {
  ...
  @$\blacktriangleright$@char[] buf = new char[1024];
  ...}
\end{lstlisting}
%\vspace{-0.3in}
%\captionof{lstlisting}{The problematic source code highlighted by \tool{} in FindBugs.}
\end{figure}

\begin{comment}
\begin{figure}
\begin{minipage}{\linewidth}
\tiny
 \begin{Verbatim}[commandchars=\\\{\}]
---------------------------------------------------------------------------------
FindBugs2.main(FindBugs2.java:1200)
 FindBugs.runMain(FindBugs.java:402)
  FindBugs2.execute(FindBugs2.java:283 )
   FindBugs2.analyzeApplication(FindBugs2.java:1089)
    DetectorToDetector2Adapter.visitClass(DetectorToDetector2Adapter.java:76)
     detect.LoadOfKnownNullValue.visitClassContext(LoadOfKnownNullValue.java:62)
      \textcolor{red}{detect.LoadOfKnownNullValue.analyzeMethod(LoadOfKnownNullValue.java:120)}
*************************Accesses to the object above************************
FindBugs2.main(FindBugs2.java:1200)
 FindBugs.runMain(FindBugs.java:402)
  FindBugs2.execute(FindBugs2.java:283 )
   FindBugs2.analyzeApplication(FindBugs2.java:1089)
    DetectorToDetector2Adapter.visitClass(DetectorToDetector2Adapter.java:76)
     detect.LoadOfKnownNullValue.visitClassContext(LoadOfKnownNullValue.java:62)
      \textcolor{red}{detect.LoadOfKnownNullValue.analyzeMethod(LoadOfKnownNullValue.java:142)}
      \textcolor{red}{detect.LoadOfKnownNullValue.analyzeMethod(LoadOfKnownNullValue.java:156)}
---------------------------------------------------------------------------------
\end{Verbatim}
\end{minipage}

\caption{\tool{} reports another object (allocated at line 119 in {\tt LoadOfKnownNullValue.java}) within its full allocation context and its accesses in FindBugs.}
\label{FindBugs2 call path}
\end{figure}
\end{comment}

\begin{figure}[t]
\begin{lstlisting}[firstnumber=111,language=java,caption={The source code highlighted by \tool{} shows the problematic object {\tt sometimesGood} (allocated at line 120) suffering from poor locality.},label=FindBugs2]
private void analyzeApplication() throws InterruptedException {
  for (Detector2 detector : detectorList) {
     ...
     @$\blacktriangleright$@detector.visitClass(classDescriptor);
     //visitClass method will call to analyzeMethod method below
  }}
private void analyzeMethod(ClassContext classContext, Method method) {
  ...
  @$\blacktriangleright$@IdentityHashMap<InstructionHandle, Object> sometimesGood = new IdentityHashMap<InstructionHandle, Object>();
  ... }
\end{lstlisting}
%\vspace{-0.3in}
%\captionof{lstlisting}{The source code highlighted by \tool{} shows the problematic object {\tt sometimesGood} (allocated at line 120) suffering from poor locality.}

\end{figure}

\subsection{Renaissance 0.10: scala-stm-bench7}
scala-stm-bench7 is a Renaissance~\cite{Prokopec:2019:RBS:3314221.3314637} benchmark, which runs stmbench7 code using ScalaSTM~\cite{scalastm} for parallelism. It is written in Scala. We run scala-stm-bench7 using the default 60 repetitions. \tool{} pinpoints a problematic object {\tt \_wDispatch} as shown in Listing~\ref{scalastm}.
%Figure~\ref{scala-stm-bench7 call path} shows a problematic object allocation in its full calling context, which is at line 619 in method {\tt grow} in the {\tt AccessHistory.scala} file. Listing~\ref{scalastm} shows the problematic object is the array {\tt \_wDispatch} (line 619). 
This object accounts for 25\% of total cache misses. With further investigation, we find that the method {\tt grow} is called frequently to adjust the {\tt \_wDispatch} array capacity and create a new {\tt \_wDispatch} array. Such frequent invocation of {\tt grow} is because the initial size of {\tt \_wDispatch} array is only 8. For optimization, we increase the initial size of {\tt \_wDispatch} array to be 512, which reduces array creation and copy by 79\%. This optimization yields a $(1.12\pm0.04)\times$ speedup to the entire program.

\begin{figure}
\begin{lstlisting}[firstnumber=615,language=java,caption={\tool{} pinpoints {\tt \_wDispatch} object suffering from high cache misses in scala-stm-bench7. },label=scalastm]
private def grow() {
  _wCapacity *= 2
  if (_wCapacity > _wDispatch.length) {
    ...
    @$\blacktriangleright$@_wDispatch = new Array[Int](_wCapacity)
  }  ... }
\end{lstlisting}
\vspace{-0.3in}
%\captionof{lstlisting}{\tool{} pinpoints {\tt \_wDispatch} object suffering from high cache misses in scala-stm-bench7. }
%\label{}
%\vspace{1em}
\end{figure}

\begin{table*}[h]
\begin{center}
\includegraphics[width=0.77\textwidth]{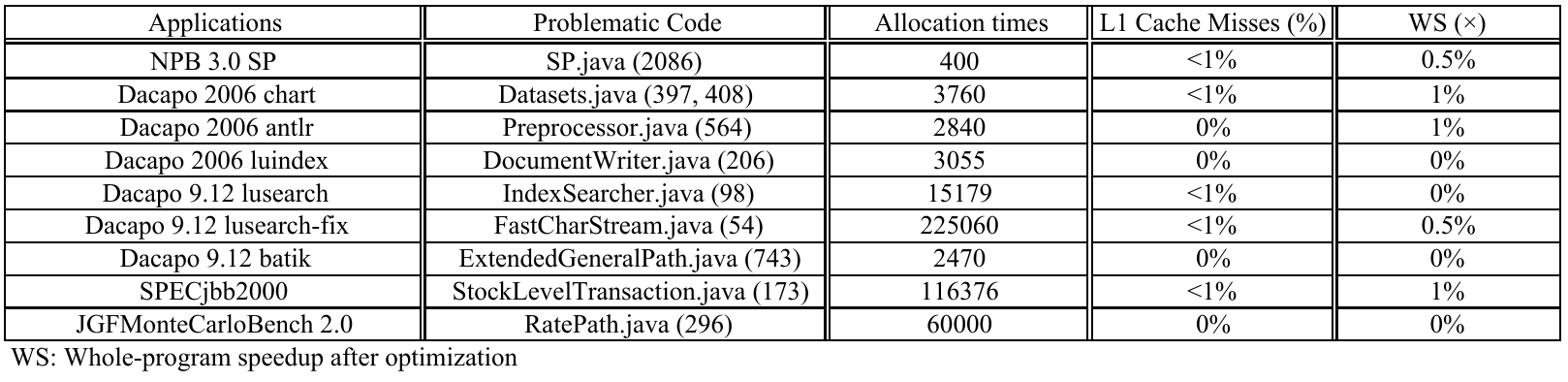} 
\end{center}
\caption{Optimizing insignificant objects yields little speedups.} 
\vspace{-1em}
\label{adversary study table}
\end{table*}

%\subsection{Issues due to Traditional Locality}
%\label{locality}
\subsection{SPECjvm2008: Scimark.fft.large}

Scimark~\cite{scimark} is a composite Java benchmark measuring the performance of numerical codes occurring in scientific applications. Scimark.fft refers to a fast Fourier transform (FFT) implementation. We run Scimark.fft with its large input released with the benchmark.
\tool{} reports the object {\tt data}, which is an array, suffering most from cache miss (accounting for 75.5\% of total cache misses). The most problematic accesses are at lines 171, 172, 174, and 175 in method {\tt transform\_internal} of class {\tt FFT}, as shown in Listing~\ref{fft}. 
From the code listing we can see that array {\tt data} is accessed in a 3-level loop nest. The innermost loop index {\tt b} increases by $2*dual$ every iteration, and $dual$ is also doubled every iteration of the outer-most loop. 
Thus, the accessing stride for {\tt data} array is large, resulting in the poor spatial locality.
For optimization, we interchange loops {\tt a} and {\tt b} to reduce the stride. This optimization reduces cache misses of the entire program by 70\%,  yielding a $(2.37\pm0.07)\times$ speedup.

\begin{figure}[t!]
\begin{lstlisting}[firstnumber=165,language=java,caption={\tool{} identifies the {\tt data} array with poor locality in SPECjvm2008: Scimark.fft.large.},label=fft]
protected void transform_internal (double data[], int direction) {
  for (int bit = 0, dual = 1; bit < logn; bit++, dual *= 2) {
    for (int a = 1; a < dual; a++) {
      for (int b = 0; b < n; b += 2 * dual) {
        int i = 2*(b + a);
        int j = 2*(b + a + dual);
        @$\blacktriangleright$@double z1_real = data[j];
        @$\blacktriangleright$@double z1_imag = data[j+1];
        ...
        @$\blacktriangleright$@data[j]   = data[i]   - wd_real;
        @$\blacktriangleright$@data[j+1] = data[i+1] - wd_imag;
        ...
}}}}
\end{lstlisting}
%\vspace{-0.3in}
%\captionof{lstlisting}{\tool{} identifies the {\tt data} array with poor locality in SPECjvm2008: Scimark.fft.large.}
%\label{fft}
\end{figure}

\subsection{Eclipse Collections}
Eclipse Collections is a comprehensive Java collections library, which enables productivity and performance by delivering an expressive and efficient set of APIs and types~\cite{Eclipse}. We run eclipse collections using CollectTest as input. After profiling eclipse collections, \tool{} reports an object that suffers from NUMA remote accesses, the Integer array {\tt result}, which is allocated at line 758 in method {\tt toArray} of class {\tt Interval} and accessed at line 245 in method {\tt batchFastListCollect} of class {\tt InternalArrayIterate} with a high percentage of NUMA remote accesses (73.4\%). By investigating the source code as shown in Listing~\ref{eclipse collections listing}, the program first calls to {\tt toArray} method to allocate and initialize an Integer array {\tt result}. And then, the program passes the {\tt result} to {\tt batchFastListCollect} method and accesses it at line 245. \tool{} detects that there's a mismatch between the allocation of {\tt result} and initialization by the master thread in one NUMA domain, and accesses by the worker threads executing in other NUMA domains. The workers all compete for memory bandwidth to access the data in the master's NUMA domain. To avoid contending for data allocated in a single NUMA domain, we optimized the program as allocating and initializing the object {\tt result} in every NUMA domain. The optimization reduces remote accesses by 41\% and increases the throughput (operations per second) by $(1.13\pm0.04)\times$.

\begin{figure}
\begin{lstlisting}[firstnumber=235,language=java]
//Interval.java
public Integer[] toArray() {
  @$\blacktriangleright$@Integer[] result = new Integer[this.size()];
  ...
  return result;
}
//InternalArrayIterate.java
private static <T> void batchFastListCollect(T[] array, ...) {
  ...
  for (int i = start; i < end; i++)
    @$\blacktriangleright$@castProcedure.value(array[i]);
  ...
}
\end{lstlisting}
\vspace{-0.3in}
\captionof{lstlisting}{\tool{} pinpoints the {\tt array} array suffering from NUMA remote accesses in Eclipse Collections.}
\label{eclipse collections listing}
\end{figure}

\subsection{Apache Druid}
\label{case2}
Apache Druid is a high performance real-time analytics database designed for workflows where fast queries and ingest matter~\cite{Druid}. We run Apache Druid using BitmapIterationBenchmark as input. After profiling SP, \tool{} reports an array an BitSet object, {\tt bitmap}, which is initialized at line 37 in constructor method {\tt WrappedImmutableBitSetBitmap} of class {\tt WrappedImmutableBitSetBitmap} and accessed at line 120 in method {\tt next} of same class. Listing~\ref{druid listing} shows the problematic object is the {\tt bitmap} (accessed at line 120), which more than half of total memory accesses are remote accesses. With further investigation, we find that the problematic object {\tt bitmap} is initialized in  constructor method {\tt WrappedImmutableBitSetBitmap} executed in one NUMA domain, but accessed by many threads in other NUMA domains. To address this problem, we parallelize the allocation and initialization for this object {\tt bitmap} to ensure that each thread first touches its own data. With this optimization, we reduce remote accesses by 47\% and increases the throughput by $(1.75\pm0.05)\times$.

\begin{figure}
\begin{lstlisting}[firstnumber=112,language=java]
public class WrappedImmutableBitSetBitmap {
  protected final BitSet bitmap;
  public WrappedImmutableBitSetBitmap(BitSet bitmap) {
    @$\blacktriangleright$@this.bitmap = bitmap;
  }
  ...
  public int next() {
    int pos = nextPos;
    @$\blacktriangleright$@nextPos = bitmap.nextSetBit(pos + 1);
    return pos;
  }
}

\end{lstlisting}
\vspace{-0.3in}
%\vspace{-1em}
\captionof{lstlisting}{\tool{} pinpoints the BitSet object {\tt bitmap} suffering from NUMA remote accesses in Apache Druid.}
\label{druid listing}
\end{figure}

\subsection{Attempts to Optimization for Insignificant Objects}
\label{adversary}
%{\color{red}Could you revisit Harry's paper to see what metrics they use to identify memory bloat? Do they just use the allocation times?}
To demonstrate the importance of PMU metrics (i.e., cache misses in our experiments) associated with the objects, we show a number of studies on attempting to optimize insignificant objects in 
Table~\ref{adversary study table}. 
All these code bases have the memory bloat problem: repeatedly allocate objects many times, and different instances have no overlap in their life intervals. 
In the table, we show the location of problematic object allocations, the number of object instances, the associated cache miss metrics, and the speedups after optimization. 
Our  studies show that these optimizations yield negligible speedups, which emphasize the fact that these objects account for very few cache misses. Thus, \tool{}'s object-centric analysis is useful to filter out insignificant objects for non-fruitful optimization.

\section{Conclusions}
\label{conclusion}
In this paper, we present \tool{}, the very first lightweight Java profiler that performs object-centric analysis to identify data locality issues in Java applications. \tool{} leverages the lightweight Java byte code instrumentation and the hardware PMU available in commodity CPU processors. \tool{} works for off-the-shelf Linux OS and Oracle Hotspot JVM, as well as unmodified Java applications. \tool{} incurs low overhead,  typically 8\% in runtime and 5\% in memory. 
These features make \tool{} applicable to the production environment. \tool{} is able to identify a number of locality issues in real-world Java applications. Such locality issues arise due to traditional spatial/temporal data locality and also due to memory bloat. 
Guided by \tool{}, we are able to perform optimization, which yields nontrivial speedups.
\tool{} is open-sourced at an \texttt{anonymous} URL.

\bibliographystyle{ACM-Reference-Format}
{\small
\bibliography{datacentric}}

\appendix

\end{document}